\documentstyle[psfig]{mn}
         
\def\ale{\mathrel{\hbox{\rlap{\hbox{\lower4pt\hbox{$\sim$}}}\hbox{$<$}}}}
\def\age{\mathrel{\hbox{\rlap{\hbox{\lower4pt\hbox{$\sim$}}}\hbox{$>$}}}}

\title[Extragalactic Correlations of Gamma-Ray Bursts] {Testing for
Extragalactic Correlations of Gamma-Ray Bursts} 

\author[Bloom, Tanvir and Wijers]{Joshua S.~Bloom, Nial~R.~Tanvir and
Ralph~A.~M.~J.~Wijers \\ Institute of Astronomy, Madingley
Road, Cambridge, CB3 0HA, England \\
email: {\tt jsbloom@ast.cam.ac.uk}}

\begin{document}

\date{MNRAS, submitted 13 May 1997}

\maketitle

\begin{abstract}

If gamma-ray bursts (GRBs) are cosmological they need not necessarily
be spatially coincident with a host galaxy---in many cosmological
models the progenitors are ejected to large distances from their
parent galaxies.  While the optical transient of GRB 970228 may be
close to a faint galaxy, there are several other galaxies in the field
which could be the host if the progenitors are ejected to $\sim 50
h^{-1}_{70}$ kpc.  If GRBs come from such a widely distributed
progenitor population then it should be possible to search for a
statistical correlation between optical transients and nearby
galaxies.  We provide a statistical framework for quantifying a
correlation between potential host galaxies and GRBs, taking into
account that galaxies are clustered.  From simulations, we estimate
that if GRBs occur within $l \ale 100$ kpc of their host, then optical
observations ($B \le 24.0$) of the galaxy field near just 6 optical
transient positions are required to find a correlation with potential
hosts at the 95 percent confidence level. The methodology constructed
herein is extendible to any cross-correlation analysis and is
especially useful if either set of objects is clustered.  Ultimately,
if the distance scale to GRBs is established as cosmological, this
analysis can be used to determine the spatial distribution of GRBs
near their host galaxies.

\end{abstract}


\section{Introduction}

After nearly thirty years since the discovery of gamma-ray bursts
(GRBs), a fading afterglow has been detected in X-ray and optical
wave-bands (Van Paradijs et al.~1997).  This observational
breakthrough has, for the first time, provided both a precise
localisation of a GRB position and broad-band spectral information.
But, while the spectral fit by Wijers, Rees \& M\'esz\'aros (1997)
does agree quite well with the cosmological blastwave theory (Rees \&
M\'esz\'aros 1992), a coherent picture of the distance scale to GRBs
has yet to emerge.  Preliminary results as to the nature of the
observed field surrounding the optical transient position have been
inconclusive: a faint extended source near the optical transient
position may have faded from images taken at Keck at two different
epochs (Metzger et al.~1997a; although see Fox et al.~1997).  The
natural assumption was that if GRBs are cosmological in origin then
the optical transient would be associated with a galaxy; if the
extension is indeed variable, then this is not the case.  Furthermore,
a spectrum of the optical transient and the extension taken with the
Keck II telescope (Tonry et al. 1997) does not reveal any tell-tale
features which would indicate the distance to GRB 970228.  So far no
extended emission has been detected from the recent burst GRB 970508
but Metzger et al.~(1997b) have reported absorption lines in the spectra
of the optical transient putting the transient beyond $z = 0.835$. If
confirmed, this would establish GRBs as cosmological.

If GRBs are cosmological, what are the other likely observational
consequences which would rule out the non-cosmological scenarios?  One
answer is that in most cosmological models GRBs will have a parent
galaxy which we can search for.  Historically this has been difficult
due to the scarcity of small error boxes: the Interplanetary Network
(IPN) of satellites yielded only a few error boxes smaller then about
$5$ arcmin$^2$ (Atteia et al.~1987). As the projected density of
galaxies is reasonably high (eg.~we expect about $30$ galaxies in 5
arcmin$^2$ boxes at $B < 24.0$) there was no lack of potential hosts
galaxies to the bursts; statistical studies of GRB positions were
thereby limited to searching for correlations on scale lengths greater
than several arcmins (Kolatt \& Piran 1996; Hurley et al.~1997).

Kolatt \& Piran 1996, for instance, do find evidence of a correlation
of bursts with Abell clusters using the crude BATSE localisations
(although see Hurley et al.~1997). Using the better, but fewer,
localisations from the IPN, there have also been claims of a bright
galaxy excess near bright bursts (Larson, McLean, \& Becklin 1996) but
this analysis uses Poissonian errors to quantify over-density,
neglecting the fact that galaxies are clustered and counts far
galaxies outside the strict limits of the IPN error boxes.  Fenimore
et al.~(1993), examining 8 of the smallest IPN error boxes, conclude
that host galaxies of GRBs must be fainter than an absolute visual
magnitude $M_v \simeq -18$.  Recent {\it Hubble Space Telescope} (HST)
observations of IPN boxes confirm this result (Schaefer et al.~1997)
and further constrain the host galaxies of GRBs to be fainter than
absolute $B_{mag} \age -17.4$. The conclusion that this requires the
hosts of GRBs to be significantly fainter than $L_{*}$ is valid only
if GRBs are standard candles and if the faintest BATSE bursts have $z
\ale 1$.  Moreover, it does not allow for the possibility that a
galaxy outside IPN error boxes could be the host to the GRBs.

With the detection of the optical transient to GRB 970228 we now have
the beginning of a set of very high precision positions and
correlation analysis should thus begin to probe smaller scales at
deeper magnitudes.  Deep HST imaging reveals faint extended emission
adjacent to the 970228 optical transient point source.  If further
observations prove that this is a distant galaxy, then it must be a
very good candidate for the host of the GRB.  However, at a magnitude
of $M_{F606W} \simeq 25.5$ (Tanvir \& Johnson 1997) the spatial
density of galaxies is high and the alignment could be a chance
coincidence.  Thus Galactic models are not ruled out and neither are
cosmological models in which the progenitors of GRBs are ejected from
their host or even only be loosely associated with the light density
from galaxies.  If GRBs arise from neutron star--neutron star (NS--NS)
coalescence, for instance, then a GRB may occur tens of kpc outside of
its host galaxy (Sigurdsson, private communication); at redshifts of
$z \sim 0.2$, such positional displacements allow an offset of the
optical transient from its host galaxy by as much as 20 arcsec. If
GRBs do occur far outside their hosts then unambiguously identifying
the parent galaxies will be difficult.  However, to rule out the
Galactic scenario, it would suffice to find a statistically
significant over-density of potential host galaxies near GRB positions
since bursts originating from within our Galaxy should be randomly
distributed with respect to underlying galaxy field.

Clearly, to distinguish between the two distance scales, there must be
a method which can both determine an over-density of potential hosts
(if there is one) and accurately estimate the significance of such an
over-density.  Ultimately, if the distance scale to GRBs is
established to be cosmological, then the same analysis will help
reveal how GRBs are distributed near galaxies.

In section 2 we provide a framework for making quantifiable
statistical statements about the degree of over-density of hosts for a
given number of GRB positions that accounts for galaxy clustering. In
section 3 we describe the simulation of a galaxy distribution which is
used in section 4 to predict the number of GRB positions needed to
find a significant correlation if GRBs are indeed associated with
distant galaxies.  In section 5 we show how the method developed
herein may be extended to other types of data and hypotheses. In other
words, should observations provide no evidence for a positive
correlation, the test developed in section 2 can be used to accept the
alternative hypothesis that GRBs {\it are not} correlated with
potential hosts to faint limits (obviously, by rejection of the null
hypothesis that there {\it is} a correlation under some given set of
assumptions).  We consider this issue in light of results from HST
images of GRB 970228.

\section{Construction of A Statistic of Over-density}

\subsection{What is a potential host?}

In an overwhelming number of viable cosmological models, a GRB
originates from near or within a galaxy.  In each model scenario,
there is a three-dimensional probability, $P_3(\vec x)$, of a GRB
originating from a given galaxy at a co-moving position, $\vec x$,
relative to its centre.  In the simplest case, we take the probability
of a GRB occurring to be constant within a sphere of radius $l$. The
projected two-dimensional probability distribution on the sky, for a
galaxy at redshift $z$ is thus:
\begin{equation}
P_2(l, \theta,z)d\theta = \frac{3 A_i(z) \sqrt{D^2 -
\theta^2}d\theta}{2\pi D^3},
\label{eq:prob2d}
\end{equation}
where
\begin{equation}
D = \frac{l H_0}{2 c} 
	\frac{(1+z)^2}{\left[ (1 + z) 
	- (1 + z)^{1/2} \right]}~{\rm radians}	
\end{equation}
in the standard, $q_o=1/2$, $\Lambda=0$, cosmology. Hereafter, the
Hubble constant is taken as $H_0=70$ km s$^{-1}$ Mpc$^{-1}$
(eg.~Tanvir et al.~1995).  For each burst, $i$, there is a total
probability normalisation, $A_i(z)$, for each galaxy in the field.  If
the results of the log $N$-log $P$ studies are to be used as a
guideline (as in Fenimore et al.~1993), then a bright GRB---such as
the expected case for most BeppoSAX X-ray detections or small IPN
error box localisations---is expected to come from a low-redshift
galaxy and thus $A_i(z)$ for that particular burst will peak at a low
redshift.

For each burst position $i$ (which is observed in a given passband,
and limiting magnitude) one constructs a probability host map,
PHM${}_i(r)$, by adding the contribution of $P_2(r,\theta,z)d\theta$
from each galaxy where $r$ is the maximum offset scale of the map. A
PHM gives the relative probability of a GRB occurring at any position
on the sky assuming a particular functional form of the offset
probability, $P_2(r,\theta,z)$.  If GRBs are not correlated with
galaxies then their positions should occur randomly on the PHM;
conversely, if GRBs are correlated with galaxies then their positions
will arise in regions of larger probability density on the PHM.

Since we do not know {\em a priori} both the intrinsic offset scale
$l$ and the functional form of the offset probability distribution, we
must assume some map scale $r$ and a form of the offset probability
distribution [$P_2(r,\theta,z)$]. In addition, construction of the PHM
requires knowledge of the redshift of each galaxy; redshifts may be
determined from photometric colours, via their observed spectra, or
simply estimated from their brightness using previous flux/redshift
studies.  Alternatively, a median redshift of galaxies in the field
may be assigned to each galaxy. In the simulations in section 3, we
show that, not surprisingly, a PHM is most sensitive to detecting
correlations if $r \simeq l$.  We also find that assigning the median
redshift to each galaxy in the map suffices in finding correlations
(see fig.~[\ref{fig:pred1}]) at least for small intrinsic offset
scales.  Note that if the map scale $r$ is sufficiently small (i.e.~if
GRBs are retained within a few kpc of a host galaxy), the PHM reduces
to a series of $\delta$-functions. In section 6 we discuss what a
reasonable value for $l$ might be for different models.

In what follows, let $\rho_{i,j}$ be the value of PHM$_i(r)$ for the
$j$th randomly selected location on map $i$; by counting the frequency
of $\rho_{i,j}$ for a large number of randomly selected points we
construct the distribution $f_i(\rho)$, which is effectively the
probability of finding $\rho$ of a potential host on the sky at any
point on the map.

\subsection{The Test for Over-densities of the True Sample}

For a given passband and limiting magnitude of an observation of each
position, there will be a corresponding number density, $\bar N_i$,
and two-point correlation function, $\omega_i(\theta)$, that
characterises the underlying distribution of galaxies from which
$f_i(\rho)$ is found. In section 3 we simulate an underlying galaxy
distribution that matches the observed properties.  In practice, it
may be better to estimate $f_i(\rho)$ directly from a number of
control fields taken at similar airmass, galactic latitude, limiting
magnitude, etc.

The normalised cumulative probability function is given as:
\begin{equation}
F_i(\rho) = \frac{\int_{\rho_{{\rm min},i}}^{\rho} f_i(\psi) d\psi}
		{\int_{\rho_{{\rm min},i}}^{\rho_{{\rm max},i}}
	 f_i(\psi) d\psi},
\label{eq:cum_prob} 
\end{equation}
where $\rho_{{\rm min},i}$ and $\rho_{{\rm max},i}$ are the extreme
values of $\rho$ on a given PHM${}_i$.  If $\rho_{i,{\rm obs}}$ is the
observed value of $\rho$ on the PHM of burst $i$ then the set of
numbers $S =\{F_i(\rho_{i,{\rm obs}} ) \}$ will be evenly distributed in the
interval [0,1] if the positions of the GRB ensemble are uncorrelated
with that of the potential host galaxy distribution.

The degree of deviation of $S$ from an evenly distributed, cumulative
set can be used to estimate the strength of the correlation between
GRBs and observed potential host galaxies.  Thus, this method allows
one to combine an inhomogeneous set of observations to test the
hypothesis that GRBs are correlated with host galaxies.

The significance of the discrepancy between the observed set $S$ and
the expected set is evaluated using the Wilcoxon statistic (see
Noether 1967 and Efron \& Petrosian 1996).  It is defined as:
\begin{equation}
W = \sqrt{12N_{\rm burst}}\left[0.5 - 
	\frac{1}{N_{\rm burst}}\sum_{i=1}^{N_{\rm burst}}
	F_i(\rho_{i,{\rm obs}})
  \right], 
\label{eq:W}
\end{equation}
and compares the average value of $S$ and the expected value of the
cumulative probability distribution ($\equiv 0.5$).  {\it The
distribution of $W$ values, $f(W)$, for an uncorrelated set of GRB
positions and potential hosts should be normally distributed with
variance unity centred about zero.} In section 3 we construct $f(W)$
for a given limiting magnitude and number of optical transient source
positions by a Monte Carlo simulation of many expected (uncorrelated)
sets (see fig.~[\ref{fig:w}]). We test the null hypothesis that GRB
optical transient source positions are uncorrelated with host galaxies
by comparing the stochastic realization of $W_{{\rm obs}}$ with the expected
distribution of $f(W)$.  Clearly if there is a positive correlation,
then the $W$-statistic will be negative (and the opposite is true for
a anti-correlation). In this formalism, the null hypothesis can be
rejected with a quantifiable confidence level by comparison of
$W_{{\rm obs}}$ with $f(W)$.
 
\section{Predictions}

In this section we construct simulations which will be used to assess
the power of the above analysis under different assumptions about the
true GRB distribution.  Specifically we will use these simulations
to ask how many observations would be required to show some
correlation of GRBs with galaxies in various scenarios.

\subsection{Constructing a Probability Host Map (PHM)}

To estimate the expected significances, we construct a galaxy
distribution which artificially replicates the observed galaxy number
density, two-point correlation function, and redshift distribution for
a given limiting magnitude.  Several galaxy distribution studies have
provided the necessary data to construct such a distribution but here
we limit ourselves to the correlation/redshift study of Roche et
al.~(1996).  For a given limiting magnitude $m$ in the R- and B-band,
they provide a redshift distribution of galaxies ($f_m(z)$), number
density ($\bar N_m$), and fit to a two-point correlation function
($\omega_m(\theta)$).  The observations were made at high galactic
latitude ($b \simeq 85^{\circ}$) and with little extinction ($E_B \simeq 0.15$
and $E_R \simeq 0.08$). This simulation is obviously extendible using
other galaxy distribution studies which provide similar data in other
passbands and/or limiting magnitudes.

\subsection{Simulating a PHM from galaxy spatial and redshift distribution}

As mentioned earlier, it may be possible to construct a PHM directly
from control fields taken under similar conditions as the optical
transient field.  Here, however, we generate a PHM artificially. To
generate an artificial galaxy distribution, we employ the truncated
hierarchy method (Soneria \& Peebles 1978) which has been adapted to
2-D projections (Infante 1994). For a given burst position $i$ we vary
the intrinsic scaling $\theta_0$ of the algorithm (see Infante 1994)
to fit the expected spatial distribution of underlying galaxies
(quantified by the spatial density distribution, $\bar N_i$ and the
amplitude and slope, $\alpha$, of the two-point correlation function).
For each galaxy in the distribution, we construct a probability
profile from eq.~[\ref{eq:prob2d}] replacing $l$ with $r$ and thereby
construct a PHM$(r)$.
\begin{figure*}
\centerline{a\psfig{file=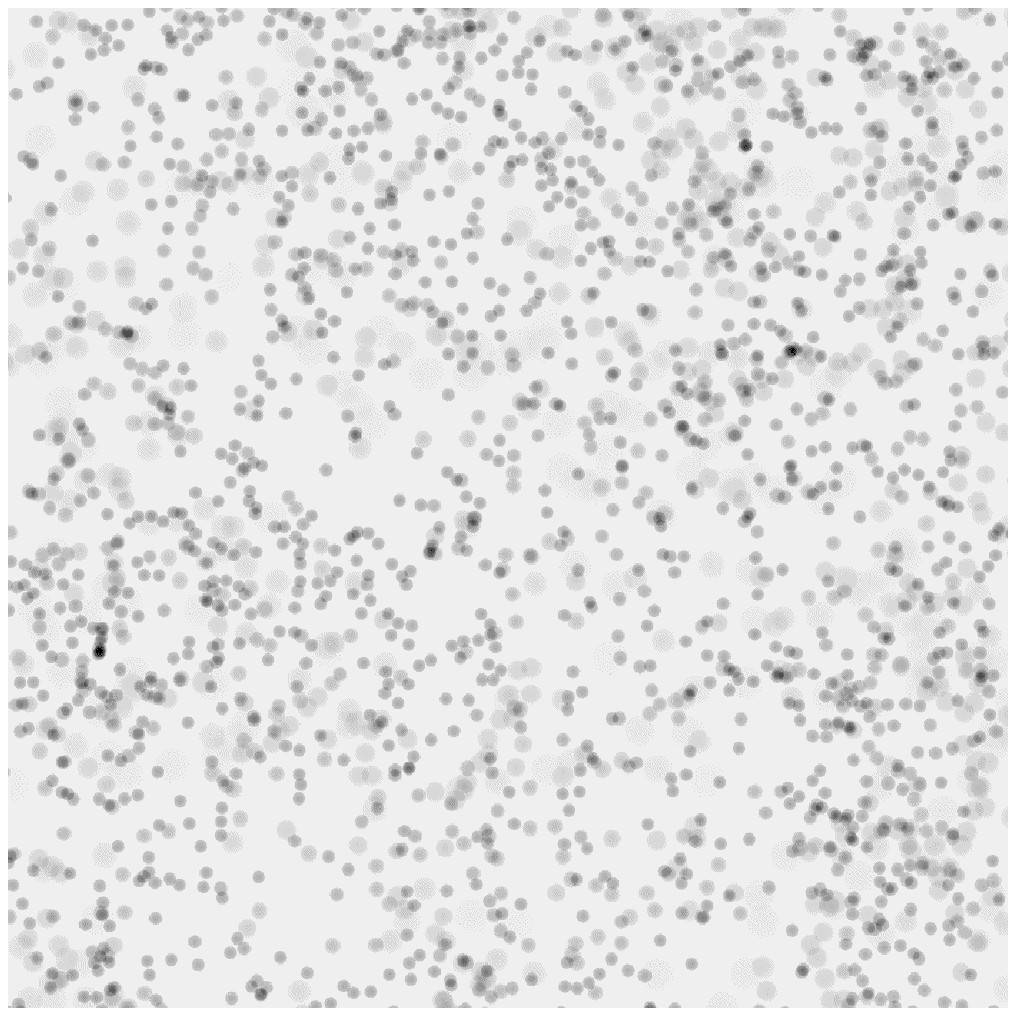,height=2.0in,width=2.0in}
\psfig{file=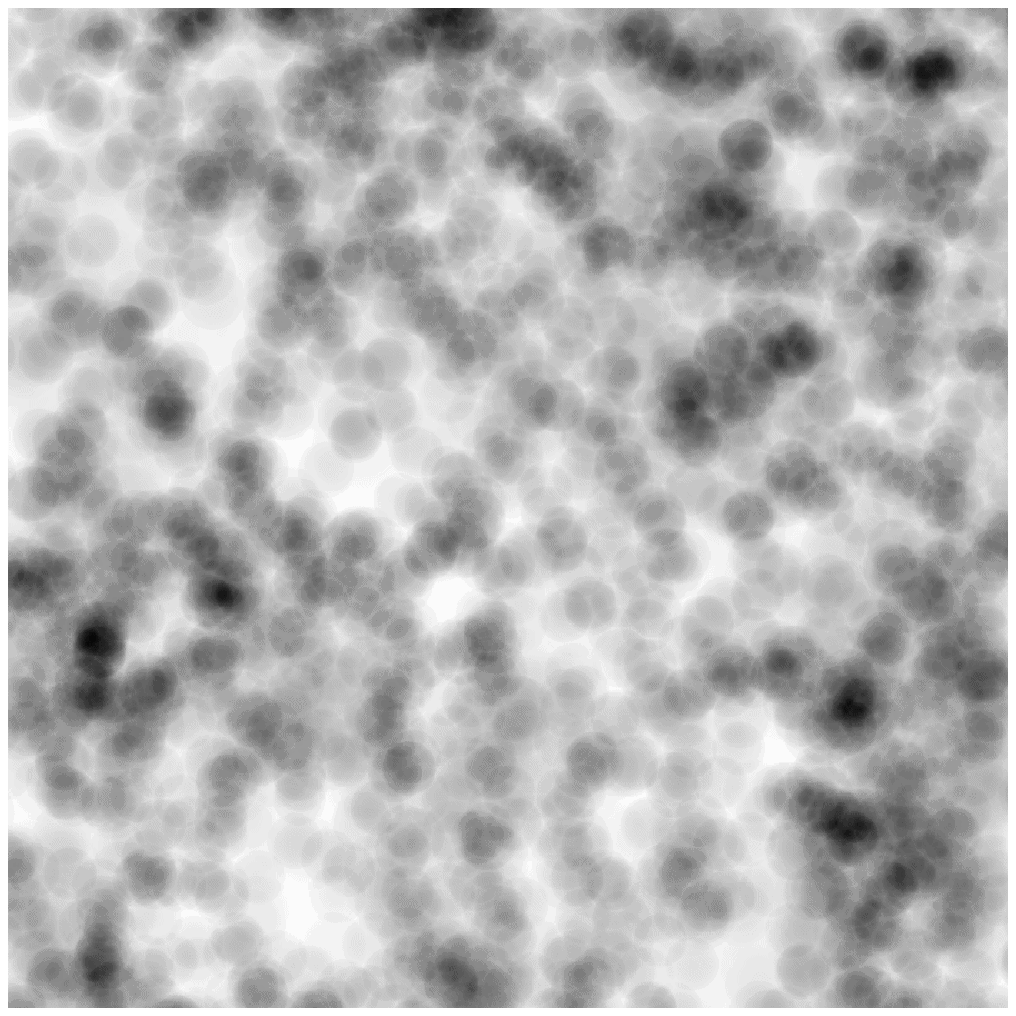,height=2.0in,width=2.0in}b
\psfig{file=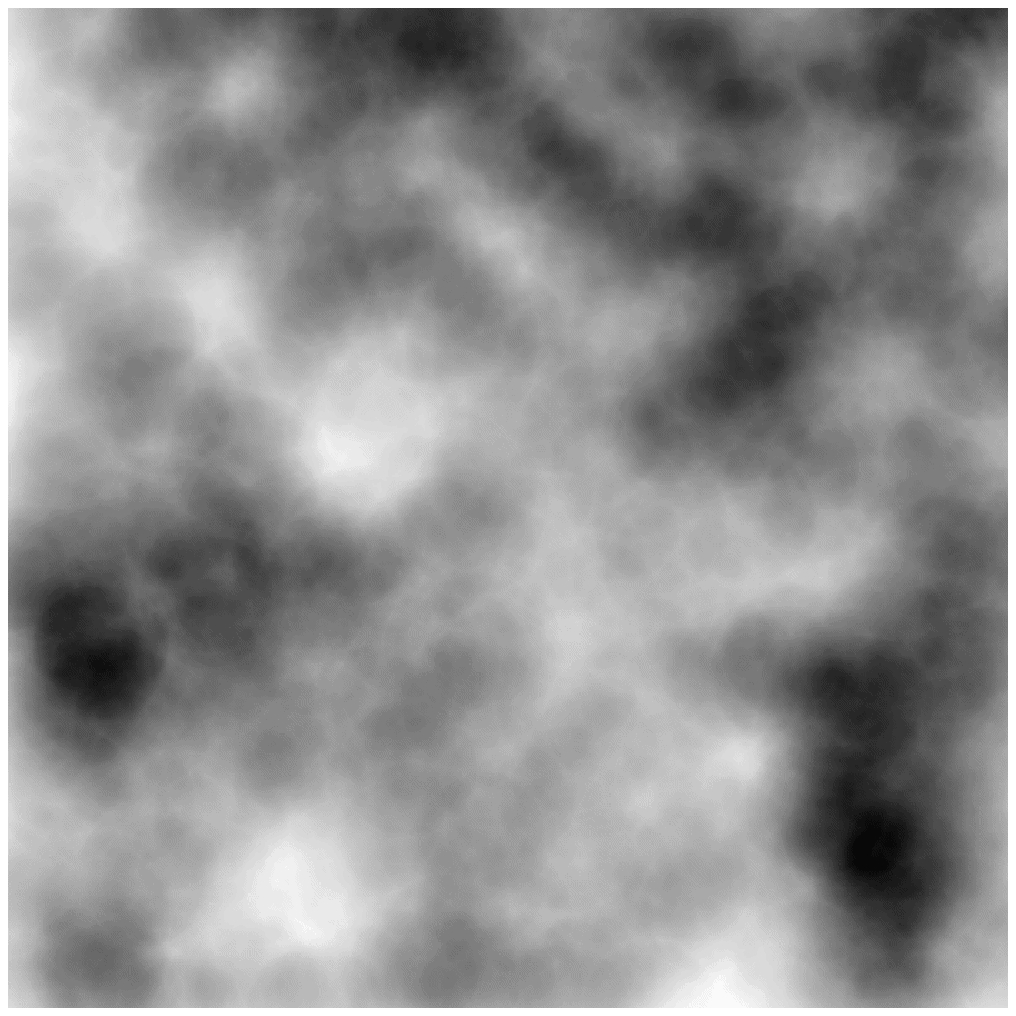,height=2.0in,width=2.0in}c}
\caption{Simulated probability host map (PHM) from galaxy fields ($B
\le 24.0$) for a) $r=50$ kpc, $r=200$ kpc, and, c) $r=500$ kpc.
Regions of higher potential host density are darker and the size of
the circular region around each galaxy is related to its redshift by
eq.~[\ref{eq:prob2d}].  Each field, comprising about 2300 galaxies, is
a 20 arcmin $\times$ 20 arcmin portion of a large PHM.  The simulated
clustering of galaxies evident gives a larger spread in $f(\rho)$ than
that derived from a randomly distributed set of galaxy positions.}
\label{fig:PHM30}
\end{figure*} 

The construction of the galaxy distribution and PHM requires a few
assumptions.
\begin{enumerate}

\item{} We take the slope of the two-point correlation function to be
$\alpha=-0.8$ which, based on numerous galaxy distribution studies
(eg.~Roche et al.~1996), seems reasonable.  In our simulation using
the truncated hierarchy method, for $B \le 24.0$, we are able to
reproduce the two-point correlation function and number density ($\bar
N=20730$ gals/deg$^2$) that match that found in Roche et al.~1996.
Bright GRBs are expected to come from moderate redshifts. We have
therefore chosen a limiting magnitude of $B \le 24.0$ which
corresponds to a median redshift of $z \sim 0.6$.  Clearly the
analysis may be extended to fainter magnitudes in which case the
expected number of burst positions required to find a correlation will
increase.

\item{}We assign a redshift to each galaxy in the generated
distribution.  Roche et al.~(1996) provide an estimated galaxy
redshift distribution as a function of limiting magnitude.  For each
cluster in an artificially generated spatial distribution, we then
assign a similar redshift ($\pm 5$ percent) to its constituent galaxies
based on a probability distribution constructed from the Roche et
al.~(1996) data.  This accounts for both for the observed distribution
of redshifts and the fact that galaxies within the same cluster have
similar redshifts.

\item{}We take $A_i(z)$ equal to unity.  Note that {\it any}
non-uniform dependency of $A_i(z)$ upon redshift will reduce the spatial
density of potential hosts and thus increase the detectability of a
correlation.  Although the distances to GRB hosts should be in accord
with distance implied by the log $N$-log $P$, it is not clear that the
log $N$-log $P$ gives a meaningful constraint on the functional form
of $A_i(z)$ since a standard luminosity of GRBs appears to be
unconstrained (Loredo \& Wasserman 1996; Horvath, M\'esz\'aros, \&
M\'esz\'aros 1996). 
\end{enumerate}

A 20' $\times$ 20' portion of three PHMs is depicted in figure
(\ref{fig:PHM30}) for various map scale lengths, $r$.  The clustering of
galaxies is evident.

\section{Simulation Results}

\begin{figure}
\centerline{\psfig{file=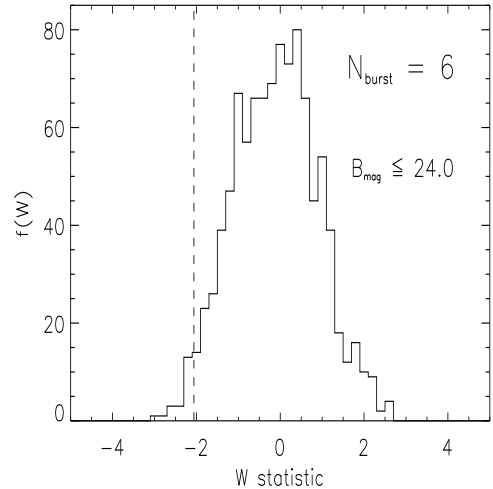,height=2.9in,width=2.9in}}
\caption{The expected distribution of $f(W)$ for an uncorrelated set
of 6 GRB positions for $l=r=100$ kpc.  Also, shown with a vertical
dashed line, is the predicted $W$ determined from a correlated set of
6 GRB positions. From the distribution, we compute that such a value
($W \simeq -2$) or smaller could occur by random chance in only
3.1 percent of cases if the GRB positions are not correlated with potential
hosts.  Thus, the null hypothesis may be rejected at greater than 95 percent
confidence.}
\label{fig:w}
\end{figure}

At a given limiting magnitude, bandpass, and number of optical
transient positions, $N_{\rm burst}$, we find the Wilcoxon
distribution $f(W)$ as prescribed in section II. We then simulate an
over-density of host galaxies near GRBs by choosing a random galaxy on
the sky and placing the optical transient position at an offset
prescribed by eq.~[\ref{eq:prob2d}].  The stochastic realization of
$W_{{\rm obs}}$ vs.~$f(W)$ is used to determine the significance of a
correlation.  Figure (\ref{fig:w}) shows an example comparison of
$W_{{\rm obs}}$ (dashed vertical line) vs.~$f(W)$.
For a given limiting magnitude, map scale $r$, and intrinsic scale
$l$, we predict the probability that the stochastic realization of $W$
is consistent with the null hypothesis,
\vskip.2cm

\centerline{$H_0$: GRB positions are uncorrelated with potential hosts.}

\begin{figure*}
\centerline{a\psfig{file=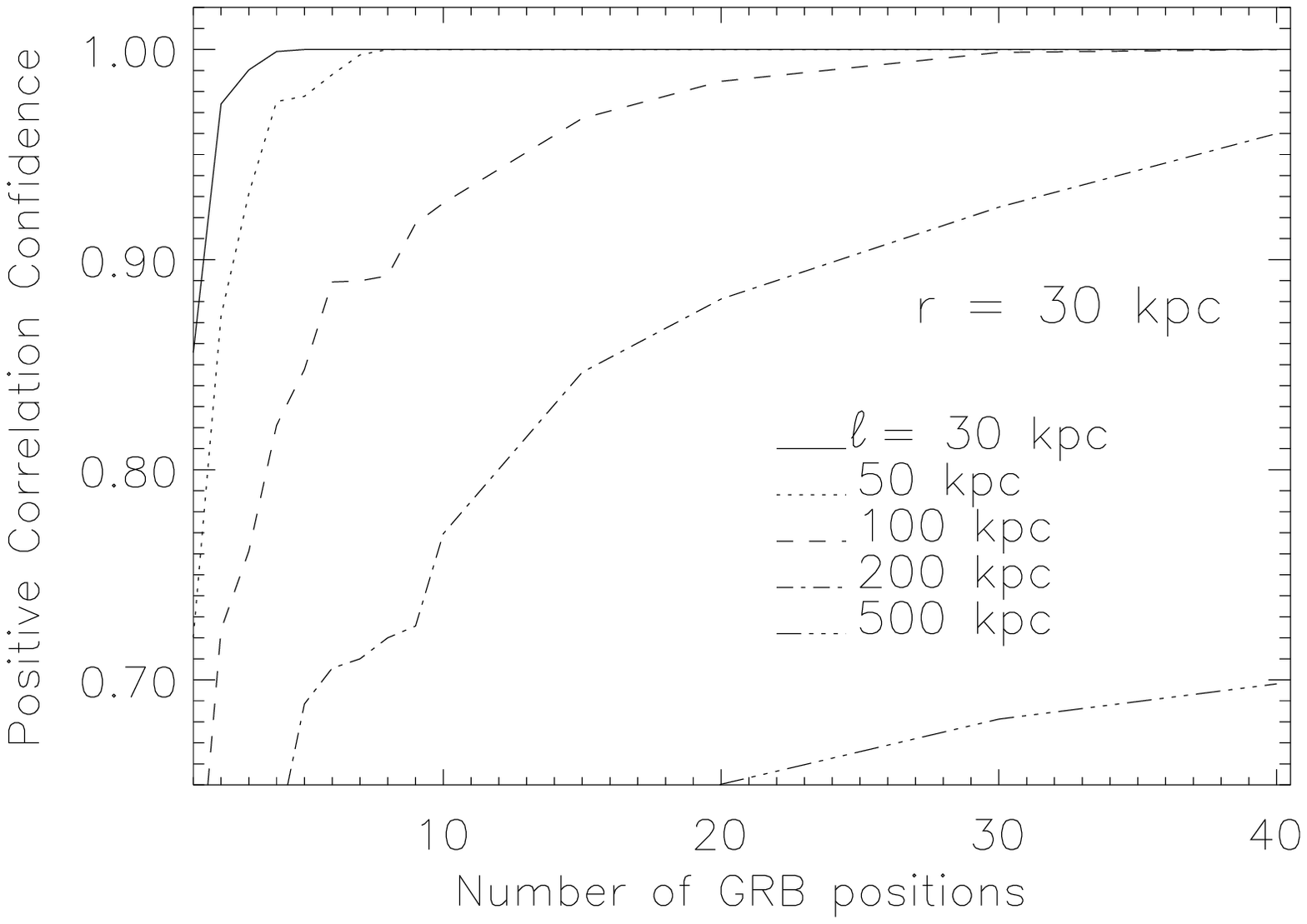,height=2.9in,width=2.9in}
	\psfig{file=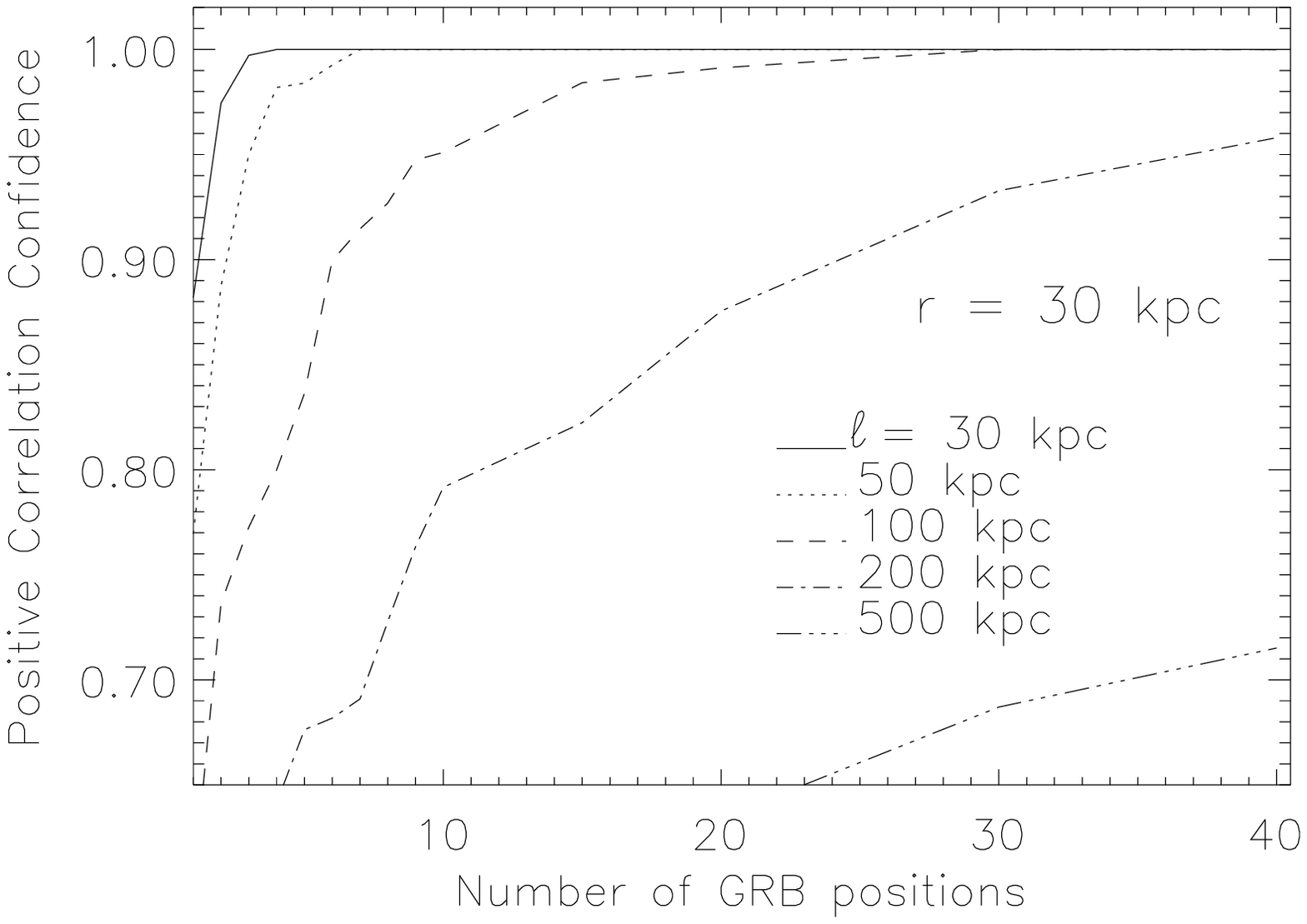,height=2.9in,width=2.9in}b}
\caption{Null hypothesis rejection confidence as a function of the
number of GRB positions. The null hypothesis tested is that GRBs are
uncorrelated with the potential host distribution.  The PHM is made
with $r=30$ kpc and taking either a) every galaxy is located at the
median redshift ($z \simeq 0.61$) of $B \le 24.0$ galaxies (Roche et
al.~1996) or b) the simulated redshift for each galaxy.  The predicted
curves for $l=30$ kpc (solid), $l=50$ kpc (dot), $l=100$ kpc (dash),
$l=200$ kpc (dot-dash), and $l=500$ kpc (dots-dash) are depicted; the
small deviations from a smooth curve are the result of a finite number
of trials in the Monte Carlo simulations.  For $l \ale 50$ kpc, only a
few burst positions are required to find a 99 percent confident correlation
of GRBs with potential hosts.  Note that with increasing values of $l$
it becomes more difficult to detect over-densities. However, other
chosen PHM scales $r$ are more sensitive to correlations at larger $l$
(see fig.~[\ref{fig:rmap}]). The predictions do not greatly differ for
each type of map.  Thus, precise knowledge of the redshifts of
galaxies will not be necessary to construct a PHM.}
\label{fig:pred1}
\end{figure*}

\begin{figure}
\centerline{\psfig{file=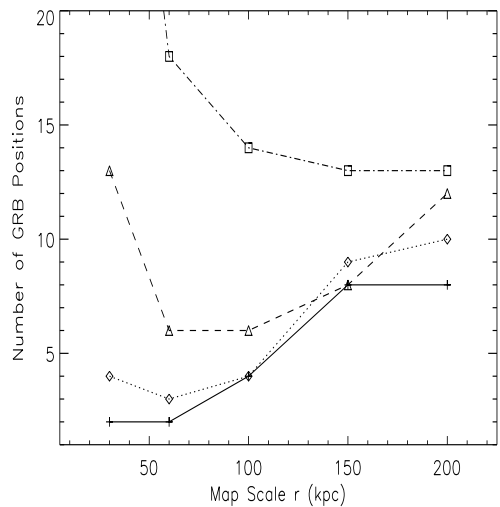,height=2.9in,width=2.9in}}
\caption{Number of GRB positions needed for 95 percent confident
correlation detection as a function of map scale $r$.  The intrinsic
scale of GRBs in the simulation are $l=30$ kpc (solid line), $l=50$
kpc (dotted), $l=100$ kpc (dashed), and $l=200$ kpc (dot--dashed).
Notice that a correlation is most readily found with $r \simeq l$.}
\label{fig:rmap}
\end{figure}

\vskip.2cm \noindent For a magnitude $B \le 24.0$ and $r=30$ kpc,
figure (\ref{fig:pred1}) shows the confidence in rejecting
$H_0$ as a function of the number of GRB positions and $l$.  Notice
that if $l$ is less than $\sim 50$ kpc, then only a few GRB positions
are required to rule out $H_0$ and conclude that GRBs are correlated
with host galaxies.  This is not a surprising result given that the
typical projected distance at $z \sim 0.3$ for $l=30$ kpc is only a
few arcsec: the expected value of PHM$(r=30)$ is close to zero and
thus any over-density is rare.
The correlation analysis is sensitive to the chosen map scale $r$ and
whether knowledge of galaxy redshifts are included.  Figure
(\ref{fig:rmap}) shows the expected number of positions needed to find
a 95 percent confident over-density (if there is one) as a function of
$l$ and $r$ where the PHMs were constructed using only galaxy
positions and the same (median) redshift ($z \simeq 0.61$) for each
galaxy.  Small values of $r$ produce PHMs that are most sensitive to a
correlation if $l$ is also small; likewise, correlations are most
readily found if $l$ is large by a choice of a large value of $r$ in
making the map.  As evidenced by the small number of bursts required
to make a significant correlation, the redshift of each galaxy,
inclusion of which strengthens the test for correlations, need not be
precisely known in making the PHM. The simulation predictions
presented in fig.~[\ref{fig:rmap}] are made from PHMs where the same
single redshift is assigned to all galaxies. Figure [\ref{fig:pred1}]
illustrates the minor strengthening of the correlation test where
knowledge of galaxy redshifts is included in construction of the PHM.

\section{Extension to other Data Sets and Hypotheses}

\subsection{Using IPN localisations}

A significant correlation of GRBs with observed galaxies may be found
without further precise locations that optical transients would
provide.  Instead, the method described above may be used on observed
fields near existing IPN localisations.  In such a case, $f_i(\rho)$
is estimated by randomly selecting positions on a PHM and taking
$\rho$ to be the average density within an error box of identical
shape and size as IPN localisation $i$.  The observed value of
$\rho_i$ is then compared against $f_i(\rho)$ as above.
 
\subsection{Are GRBs Uncorrelated with Observable Galaxies?}

\begin{table*}
\label{table:field}
\centering
\caption{Gamma-ray burst 970228 constraints on the minimum offset of
  GRBs from their host galaxy.  The data found is from cleaned HST PC1
  images taken on March 26 and April 7 (Tanvir \& Johnson 1997).
  Instrumental magnitudes are derived in two passbands and errors are
  found from systematic ($\simeq 0.1$) and photometric
  uncertainties. Zero-point calibration is taken from Holtzman et
  al.~(1995). Redshifts are estimated from an F606W magnitude
  comparison with galaxies in the Hubble deep field (Cowie 1997). The
  redshift at which 90 percent of the Cowie galaxies of similar
  magnitudes have larger photometric redshifts is denoted by $z_{90}$
  (similarly, $z_{50}$ is the median redshift of galaxies with similar
  magnitudes).  This corresponds to a constraint on the minimum offset
  $l_{90}$ (assuming $q_{\circ} = 0.5$) of GRBs from their host
  galaxies.}
\vspace{0.5cm} 
  \begin{tabular}{lcccccccl}	
\hline
\multicolumn{1}{c}{Galaxy} &\multicolumn{2}{c}{Position
  (J2000)}&$D$&\multicolumn{2}{c}{Magnitude}&\multicolumn{1}{c}{Redshift}& $l_{50}$ ($l_{90}$) 
  &Galaxy Property \\
 Number  & RA (hms)  &  DEC (deg)              & (arsec)  & F606W
& F606W $-$ F814W &  \multicolumn{1}{c}{$z_{50}$ ($z_{90}$)} & ($h_{70}^{-1}$ kpc) \\
\hline
1 & 5:01:46.50 & 11:46:50.2 & 4.1 & $26.3\pm 0.4$ & 0.9  & 1.34 (0.25)&
  25 (14)   \cr
2 & 5:01:46.20 & 11:46:45.4 & 10.8 & $21.8\pm 0.1$& 1.1  & 0.31 (0.09)&
	43 (17) & spiral, star-forming?  \cr
3 & 5:01:46.56 & 11:46:46.1 & 7.7 & $23.7\pm 0.1$ & 0.8  & 0.70 (0.20)&
	44 (23) \cr
4 & 5:01:47.17 & 11:46:49.3 & 8.8  & $25.5\pm 0.21$& 0.8 & 0.88 (0.62)&
	53 (48) \cr
5 &  5:01:47.15 & 11:46:54.4 & 5.2 & $25.2\pm 0.23$& 1.0 & 0.87 (0.25) &
	31 (18) & emission-line galaxy \cr		
  &             &             &      &                &&$z=0.638^{a}$&
	28.9 \cr 
6 &  5:01:46.94 & 11:47:04.1 & 11.4 &$25.4\pm 0.30$& 0.5 &0.84 (0.61)&
	67 (62)    \cr
\hline
  \end{tabular} 

\raggedright $^a$ Spectroscopic redshift from Tonry et al.~1997.
\end{table*}

If the extension near the position of GRB 970228 is not a galaxy there
are no hosts at the faint HST limit ($M_{F606W} \ale 26.5$) for a
GRB/host-galaxy offset of $l \ale 14 h_{70}^{-1}$ kpc (see table 1).
If we take the less faint limits of $B \le 24.0$ commensurate with
simulations, then the somewhat stronger constraint of $l_{{\rm min}}
\age 17 h_{70}^{-1}$ kpc comes from galaxy 2 in table 1.  As the
redshifts of the galaxies in the HST fields are yet to be determined
spectroscopically, these offset lower limits should be considered
approximate.

While this one optical transient position does not provide a strong
constraint, it does seem that either it has a surprisingly faint host
or GRBs occur at large offsets from their host galaxy.  If the
extension is indeed a galaxy, then it is so faint that there is a
non-negligible chance (few percent) that the optical transient
position is coincident with it by random chance.

The absence of a detectable galaxy down to faint magnitudes
($M_{F606W} \ale 25$) near the optical transient position of GRB
970228 gives sufficient impetus to the formulation of the alternative
null-hypothesis that GRB positions are correlated with potential
hosts.  A significant rejection of this hypothesis will place
important constraints on most cosmological models.  Within such a
hypothesis, $f(\rho)$ is constructed by choosing positions near
galaxies (rather than at random) from a PHM.  By examination of the
HST fields taken on 26 March and 7 April, already one can rule out
$H_0$ for $l \ale 17 h_{70}^{-1}$ kpc for $B \ale 24$ (galaxy 2; table
1). Obviously, GRBs could originate from low-surface brightness
galaxies or at large redshifts, in which case the detectability of a
correlation would be difficult.

\section{Discussion}

Regardless of the interpretation of the emission near the optical
transient position, it is clear that GRB 970228 did not originate from
a bright, nearby galaxy (see table 1) confirming the systematic
findings of Schaefer et al.~(1997) and Fenimore et al.~(1993).  It
should be emphasised that the ``no-host'' conclusion of Schaefer et
al.~(1997) and Fenimore et al.~(1993) is based strongly on the
assumption that the GRB population follows a non-evolving, standard
candle luminosity in as much as it is argued that bright bursts should
come from nearby galaxies.  But from an observational standpoint,
Loredo \& Wasserman (1996) have concluded that the width of the
luminosity function of GRBs is poorly constrained and Horvath,
M\'esz\'aros, \& M\'esz\'aros (1996) find that a large range of
luminosities acceptably fit the log $N$-log $P$ distribution.
Moreover, theories of cosmological shocks and blastwaves predict a GRB
luminosity which can vary greatly from burst to burst (eg.~Fenimore,
Madras, \& Nayakskin 1996); instead total energy is more likely to be
standard (see Bloom, Fenimore, \& in 't Zand 1997).  Thus, robust
conclusions about (non-)correlations of hosts with GRBs should not
rely on the above assumptions.

The specific correlation hypothesis that is tested is wholly
determined by the functional form and dependences in
eq.~[\ref{eq:prob2d}].  The basic hypothesis we have tested in these
simulations is that ``the progenitors of GRBs originate (and are
possibly ejected) from observable galaxies.'' If mass in the Universe
scales with light, this can also be seen as a test that the rate
density of GRBs is highest at the peaks of the matter distribution.
Of course while this hypothesis is somewhat model dependent, it
effectively introduces only one free parameter, namely $l$ (and to a
lesser extent $P_2$).

We conservatively estimate the maximum intrinsic offset scale to be of
order $l=400$ kpc.  This corresponds to a constant progenitor outward
velocity $v_{{\rm pro}} = 400$ km/s and a collision time scale of
$\tau = 10^{9}$ years.  If all GRBs occur at such large distances from
their host galaxy, it would be difficult to find a significant
correlation with a small collection of GRB positions.  However,
simulations have shown for NS-NS models that $l$ may be an order of
magnitude smaller and $P_2(l)$ is likely to be very peaked near the
centre for most galaxies, so that the effective $l$ might be
significantly lower than 400 kpc (Lipunov et al.~1995).  Therefore,
whatever the true $l$ is, we believe that the assumed $P_2(r=l)$, is
quite conservative by over-predicting the true extension of GRBs from
their hosts.

Since $l$ is not known it is impossible to know {\it a priori} which
map scale $r$ will optimise the detection of a correlation.  The
predictions of a 95 percent confident correlation after about six
optical transient detections is based on a fixed $r=60$ kpc and
assuming $l \le 100$ kpc (see fig.~[4]). Although a detected
correlation thusly will be adequate for resolving the distance scale
to GRBs debate, ideally one would like to determine $l$ in addition.
As a correlation search over a range of map scales $r$ is likely to
increase the random chance of finding a correlation, it will be
necessary to determine $l$ {\it a posteriori} via Bayesian analysis.

One limitation of the statistical method presented is that the angular
size of the PHM must be significantly larger than the projected
angular offset $D$ of GRBs from their host.  As most cosmological
models would have GRBs originate at distances significantly less than
1 Mpc from their host galaxy, this corresponds to a minimum PHM size
of about $30$ arcmin $\times$ 30 arcmin.  The correlation analysis
hinges on an accurate modelling of the expected probability density
distribution ($f(\rho)$) and thus control fields must be selected
quite carefully in order not to bias the PHM.  We will examine the
salient issues of estimating $f(\rho)$ in a future paper, which
presents the results of a deep optical IPN error box correlation
analysis.

\section{Conclusions}

We have presented a formalism which may be used to quantify the degree
of over-density of potential host galaxies near GRB positions; if GRBs
are found to be cosmological in origin this formalism may also be used
to determine the intrinsic offset of GRBs from their hosts.  The broad
hypothesis we have chosen to address is that GRBs originate within $l$
kpc of their host galaxy.  In our simulations, we find that for $l
\ale 100$ kpc, encompassing most predictions of current cosmological
models, a 95 percent significant correlation may be found with about
six positions. This prediction corresponds to simulations of galaxy
fields of $B \le 24.0$ where the median redshift of galaxies is $z
\sim 0.6$.  If GRBs are correlated with galaxies at comparable
redshifts, such correlation may found within a year if BeppoSAX
detections lead us to a handful of optical transient localisations.

However, the extension near the optical transient of GRB 970228 is
very faint ($M_{F606W} \simeq 25.5$ implying a redshift $z \sim 1$),
which suggests that GRBs may only correlate with very distant
galaxies.  In this case, correlations will be detectable only if $l$
is sufficiently small.  If the fading extension near the optical
transient of GRB 970228 is not a galaxy, we find that the progenitor
system must have travelled a minimum distance $l_{{\rm min}} \age 14
h_{70}^{-1}$ kpc from its host galaxy before producing a GRB.
Assuming that galaxy detection in the HST exposures is complete to
distances of the bright burst, this effectively requires GRBs to
originate well outside the nucleus of their host galaxies. In any
event, the absence of a bright host galaxy indicates that the
non-evolving standard luminosity derived from the log $N$-log $P$
distribution inadequately describes the true distribution of
cosmological GRBs.

\section*{Acknowledgements}

The authors would like to thank S.~Sigurdsson, M.~Rees, E.~Fenimore,
S.~Phinney, O.~Lahav, O.~Almani, and E.~Blackman for helpful
discussions and insightful comments. We wish to acknowledge the
support of the Herchel Smith Harvard Scholarship (JSB) and support
from a PPARC postdoctoral fellowship (RAMJW).

\end{document}